\documentclass[aps,twocolumn,prl,showpacs,balancelastpage,amssymb,groupedaddress]{revtex4}
\usepackage[latin1]{inputenc}
\usepackage[T1]{fontenc}
\usepackage[english]{babel}
\usepackage{graphicx}
\usepackage{amssymb}
\usepackage{amsmath}
\usepackage{amscd}
\usepackage{eucal}
\usepackage{color}
\usepackage{bm}
\newcommand{\bi}{\bibitem}
\def\det{\delta }

\usepackage{epstopdf}
\begin{document}
\title{Quantum Reversibility and Echoes in Interacting Systems}
\author{C.~Petitjean$^1$ and Ph.~Jacquod$^2$}
\affiliation{$^1$ D\'epartement de Physique Th\'eorique,
Universit\'e de Gen\`eve, CH-1211 Gen\`eve 4, Switzerland \\
$^2$ Physics Department, 
   University of Arizona, 1118 E. 4$^{\rm th}$ Street, Tucson, AZ 85721, USA}
\date{\today}
\begin{abstract}
In Echo experiments, imperfect 
time-reversal operations are performed on a subset
of the total number of degrees of freedom. To capture the physics of these
experiments, we introduce a partial fidelity ${\cal M}_{\rm B}(t)$,
the Boltzmann echo, where only
part of the system's degrees of freedom can be time-reversed. 
We present a semiclassical calculation of ${\cal M}_{\rm B}(t)$. 
We show that, as
the time-reversal operation is performed more and more accurately, 
the decay rate of ${\cal M}_{\rm B}(t)$ saturates at a value given by
the decoherence rate of the controlled degrees of freedom due to their coupling
to uncontrolled ones. We connect these results with NMR spin echo experiments.
\end{abstract}
\pacs{05.45.Mt,03.65.Ud,05.70.Ln,03.67.-a\\[-5mm]}
\maketitle 

One of the central problems faced by the founders of statistical physics in
the last decades of the nineteenth century was to reconcile the 
time-asymmetric evolution of macroscopic systems with time-symmetric
microscopic dynamics \cite{loschmidt}. They came up with a probabilistic
solution to this {\it irreversibility paradox}.
Macroscopic states, they argued, are superpositions of an enormous amount
of microscopic states, the majority of them evolving in accordance with the
second law of thermodynamics. 
The likelihood that a macroscopic state violates the second law
of thermodynamics is thus minute, typically exponentially small in the number
of atoms it contains. Irreversibility at the macroscopic level
follows ``by assuming a very improbable (i.e. with a very low entropy)
initial state of the entire universe'' \cite{lebowitz,boltzmann}.
This mechanism works equally well in either quantum or classical 
systems. 

Simple mechanisms of irreversibility already exist at the microscopic level 
in chaotic (in particular mixing) 
classical systems with few degrees of freedom. As a matter of fact,
mixing ensures that, after a sufficiently long evolution time,
two initially well separated phase-space distributions will evenly fill
phase-space cells of any given size. Since phase-space points can
never be located with infinite precision, irreversibility sets in
after mixing has occurred on a scale smaller than the 
phase-space resolution scale. This mechanism cannot be carried
over to quantum systems, however, mostly 
because the Schr\"odinger time-evolution is unitary, in either real- or
momentum-space.
Microscopic quantum systems are generically stable under
time-reversal, even when their classical counterpart is irreversible 
\cite{dima}. Peres instead suggested to investigate quantum irreversibility 
at the microscopic level through the fidelity
\begin{equation}
{\cal M}_{\rm L}(t)=\left\vert\left \langle \psi_0 \left\vert 
\exp[i H t] \exp[-i H_0 t]
\right \vert  \psi_0 \right\rangle\right \vert^2,
\end{equation}    
with which a quantum state $ \psi_0$ can be reconstructed by 
inverting the dynamics after a time $t$ with a perturbed Hamiltonian 
$H =H_0 +\Sigma $ \cite{peres}. Because of its connection
with the gedanken time-reversal experiment proposed by Loschmidt in his
argument against Boltzman's H-theorem \cite{loschmidt},
${\cal M}_{\rm L}(t)$ has been dubbed the {\it Loschmidt Echo} by
Jalabert and Pastawski \cite{Jal01}.

Echo experiments abound in 
nuclear magnetic resonance \cite{hahn,nmr}, optics \cite{kurnit}, 
atomic \cite{davidson}, and condensed matter physics \cite{nakamura}.
Fundamentally, they are all based on the same principle of a sequence
of electromagnetic 
pulses whose purpose it is to reverse the sign of the Hamiltonian,
$H_0 \rightarrow -H_0$, by means of effective changes of coordinate 
axes \cite{hahn}. Imperfections in the pulse sequence result instead in
$H_0 \rightarrow -H_0-\Sigma$, and one therefore 
expects the Loschmidt Echo 
to capture the physics of the experiments. This line of reasoning 
however neglects the fact that the time-reversal operation affects at best
only part of the system, for instance because the system is composed of 
so many degrees of freedom, that the time arrow can be inverted only for
a fraction of them. This is generically the case, as any
system is coupled to an external, uncontrolled environment.
To capture the physics of echo experiments one thus
has to take into account 
that (i) the system decomposes into two
interacting subsystems 1 and 2; (ii) the initial state of the controlled
subsystem 1 is prepared, i.e. well defined, and its final state is measured 
and compared to the initial one; (iii) both the initial and final states of 
the uncontrolled subsystem 2 are unknown; (iv) the
Hamiltonian of system 1 is time-reversed with some tunable accuracy, however
both the Hamiltonian of system 2 and the interaction between the two
subsystems are uncontrolled. We therefore
propose to investigate the physics of echo experiments by means of 
the following partial fidelity (we set $\hbar \equiv 1$)
\begin{eqnarray}\label{irrevtest}
{\cal M}_{\rm B} (t) = \Big \langle \big\langle \psi_1  \big|    
{\rm Tr}_2 \left[
e^{-{\it i }{\cal H}_{\rm b} t } e^{-{\it i}{\cal H}_{\rm f} t } \rho_0 
e^{ {\it i }{\cal H}_{\rm f} t } e^{ {\it i}{\cal H}_{\rm b} t }
\right]\big| \psi_1 \big\rangle \Big \rangle,
\end{eqnarray}
where the forward and backward (partially time-reversed) Hamiltonians 
read 
\begin{subequations}\label{hamiltonians}
\begin{eqnarray}
{\cal H}_{\rm f} & = & H_1 \otimes I_2 + I_1 \otimes H_2 + {\cal U}_{\rm f}, \\
{\cal H}_{\rm b} & = & -[H_1+\Sigma_1] \otimes I_2 
+ I_1 \otimes [H_2+\Sigma_2] + {\cal U}_{\rm b}.
\end{eqnarray}
\end{subequations}
The experiment starts with an initial density matrix 
$\rho_0  =  |\psi_1\rangle\langle \psi_1| \otimes \rho_2$,
which is propagated forward in time with ${\cal H}_{\rm f}$. 
After a time  $t$, 
we invert the dynamics of system 1. The imperfection in that time-reversal
operation is modelled by $\Sigma_1$, while $\Sigma_2$ allows for 
system 2 to be affected by this operation (we will see below that tracing
over the degrees of freedom of 
system 2 makes ${\cal M}_{\rm B}(t)$ independent of either $H_2$ or
$\Sigma_2$). We leave open the possibility 
that the interaction between the two systems is affected
by the time-reversal operation, i.e. ${\cal U}_{\rm f}$ may or may not
be equal to ${\cal U}_{\rm b}$. Because one has no control over system 2,
the corresponding degrees of freedom are traced out. For the same reason,
the outmost brackets in Eq.~(\ref{irrevtest}) indicate an average over
$\rho_2$.
We name ${\cal M}_{\rm B}(t)$ the {\it Boltzmann echo} to stress its 
connection to Boltzmann's
counterargument to Loschmidt that time cannot be  
inverted for all components of a system with many degrees of freedom. 

In this article, we present a semiclassical 
calculation of the Boltzmann echo for two classically chaotic 
subsystems along the lines of Refs.\cite{Jal01,Jac04,Pet05}, and
compare our results with those obtained from a Random Matrix Theory (RMT)
treatment of the problem. 
Our main result is that, in the regime of classically weak
but quantum mechanically strong
imperfection $\Sigma_1$ and coupling ${\cal U}_{\rm f,b}$, 
${\cal M}_{\rm B}(t)$ is the sum of two exponentials
\begin{eqnarray}\label{eq:LEmod} 
{\cal M}_{\rm B}(t) \simeq \exp\left[-  \left(  \Gamma_{\Sigma_1}+ 
\Gamma_{\rm f} +
\Gamma_{\rm b} \right) t \right]+  
\alpha_1  \exp\left[-\lambda_1 t\right].
\end{eqnarray}
Here, $\alpha_1$ is a weakly time-dependent prefactor,
 $\lambda_1$ is the classical Lyapunov exponent of system 1, and
$\Gamma_{\Sigma_1}$ and $\Gamma_{\rm f,b}$
are given by classical correlators for $\Sigma_1$ and ${\cal U}_{\rm f,b}$ 
respectively (see below). Equivalently, they can be regarded as the golden 
rule width of the Lorentzian
broadening of the levels of $H_1$ induced by $\Sigma_1$ and
${\cal U}_{\rm f,b}$ respectively
\cite{Jac01}. Together with the one-  and two-particle
level spacings $\delta_{1,2}$ and bandwidths $B_{1,2}$, they define
the range of validity of the semiclassical approach as
$\delta_1 < \Gamma_{\Sigma_1}<B_1$, $\delta_2 < \Gamma_{\rm f,b}< B_2$ 
\cite{Jac01,Jac04,Pet05}. 
The second term on the right-hand side of Eq.~(\ref{eq:LEmod}) 
exists exclusively for a
classically meaningful initial state $\psi_1$ such as a Gaussian wavepacket
or a position state, but the first term is much more generic. It emerges
from both a semiclassical or a RMT treatment and does not depend on the
initial preparation $\psi_1$ of system 1. Other regimes of decay exist,
which we here mention for the sake of completeness.
For quantum mechanically weak $\Gamma_{\Sigma_1} \ll \delta_1$ and
$\Gamma_{\rm f,b} \ll \delta_2$,
one has a Gaussian decay,
\begin{equation}\label{gaussiand}
{\cal M}_{\rm B}(t)= \exp\left[-\left( \overline{\Sigma_1^2}/4+
\overline{ {\cal U}_{\rm f}^2} /2 + \overline{ {\cal U}_{\rm b}^2 }/2  \right) t^2\right],
\end{equation}
in term of
the typical squared matrix elements 
of $\Sigma_1$ and ${\cal U}_{\rm f,b}$.
Also, at short times a parabolic decay
of ${\cal M}_{\rm B}(t)$ prevails for any coupling strength.
Finally, if system 1 is integrable, the decay of ${\cal M}_{\rm B}(t)$
is power-law in time.

The equivalence between Boltzmann and Loschmidt echoes is broken by
$\Gamma_{\rm f,b}$, the decoherence rate of system 1 induced by the coupling
to system 2 (or by $\overline{ {\cal U}_{\rm f,b}^2}$ at weak interaction). 
Skillfull experimentalists can thus investigate decoherence
in echo experiments with weak time-reversal imperfection $\Sigma_1$ for
which $\Gamma_{\Sigma_1} \ll \Gamma_{\rm f,b}$, and thus
${\cal M}_{\rm B}(t) \simeq
\exp[-(\Gamma_{\rm f}+\Gamma_{\rm b}) t]$ (or 
${\cal M}_{\rm B}(t) \simeq \exp[-(
\overline{ {\cal U}_{\rm f}^2}+ \overline{ {\cal U}_{\rm b}^2 }
) \; t^2/2] $ at weak interaction)
as $\Sigma_1$ is reduced. 
This might well be the explanation
for the experimentally observed $\Sigma_1$-independent decay of polarization 
echoes \cite{levstein}.

We now present our calculation. As starting point, we take 
chaotic one-particle Hamiltonians $H_{1,2}$, and a smooth 
interaction potential 
${\cal U}$ which depends only on the distance between the particles.
We assume that it is characterized by a typical classical length scale, which 
in particular is larger than the de 
Broglie wavelength $\sigma$ of particle 1. For pedagogical reasons, we take
narrow Gaussian wavepackets for the initial state of both particles,
$\psi_i({\bf q})= \langle{\bf q}\vert \psi_{i} \rangle = 
(\pi \sigma^2)^{-d_i/4} \exp[{\it i}{\bf p}_i \cdot 
({\bf q}-{\bf r}_{i})-|{\bf q}-{\bf r}_{i}|^2/2 \sigma^2]$. 
We note however that within our semiclassical approach,
more general states can be taken for
the uncontrolled system 2, such as random pure states 
$\rho_2 =\sum_{\alpha\beta} a_{\alpha} 
a_{\beta}^{\ast}  \vert \phi_{\alpha} \rangle \langle \phi_\beta \vert$,
random mixtures  
$\rho_2 =\sum_{\alpha} |a_{\alpha}|^2 
\vert \phi_{\alpha} \rangle \langle \phi_\alpha \vert$ or 
thermal mixtures $\rho_2 =\sum_n 
\exp\left[-\beta E_n\right]\vert n\rangle \langle n \vert$.
Arbitrary initial states for both subsystems 
can be considered within the RMT approach.

From Eqs.~(\ref{irrevtest}) and (\ref{hamiltonians}) we can 
rewrite ${\cal M}_{\rm B}(t)$ as \\[-16mm]
\begin{widetext}
  \begin{eqnarray}
{\cal M}_{\rm B}(t) &=& \int {\rm d}{\bf z}_2 \;\;\;
\Bigg |
\int \prod_{i=1}^2 {\rm d} {\bf x}_i  \prod_{j=1}^{3} {\rm d} {\bf q}_j \;
\psi_1({\bf q}_1)\psi_2({\bf q}_2)\psi^{\dagger}_1({\bf q}_3) 
 \left\langle {\bf q}_3, {\bf z}_2 \left\vert e^{-{\it i }{\cal H}_{\rm b} t }
\right\vert {\bf x}_1, {\bf x}_2  \right\rangle
\left\langle{\bf x}_1, {\bf x}_2  \left\vert e^{-{\it i }{\cal H}_{\rm f}  t} 
\right\vert {\bf q}_1, {\bf q}_2 \right\rangle
\Bigg |^2 .
\end{eqnarray}
We next introduce the semiclassical propagators
($a={\rm f,b}$ labels forward or backward 
evolution; $ \epsilon^{(f)} = - \epsilon^{(b)} = 1$),
\begin{eqnarray}
 \left\langle  {\bf x}_1, {\bf x}_2 \left\vert e^{-{\it i}{\cal H}_{a} 
t} \right \vert {\bf q}_1, {\bf q}_2 \right\rangle \; =  \sum_{s_1,\,s_2} {\cal C}_{s_1,s_2}^{1/2} 
\exp\left[{\it i} \left\{  \epsilon^{(a)}  S^{(a)}_{s_1}({\bf q}_1,{\bf x}_1;t) + 
S^{(a)}_{s_2}({\bf q}_2,{\bf x}_2;t) + {\cal S}^{(a)}_{s_1,s_{2}}({\bf q}_1,{\bf x}_{1};
{\bf q}_{2},{\bf x}_{2};t)\right \} \right],
 \label{SCproF} 
\end{eqnarray}
which are expressed as sums over pairs of classical trajectories, 
labeled $s_i$ ($l_i$) for particle $i$ connecting ${\bf q}_i$ to 
${\bf x}_i$ in the time $t$ with dynamics determined by 
$H_i$ or $H_i+\Sigma_i$. Under our assumption of a classically weak
coupling, classical trajectories are only determined by the 
one-particle Hamiltonians. 
Each pair of paths gives a 
contribution containing one-particle action integrals denoted by $S_{s_i}$
(where we included the Maslov indices) and two-particle action 
integrals  ${\cal S}^{({\rm f,b})}_{s_1,s_2}=\int_0^{t} {\rm d}\tau 
{\cal U}_{{\rm f,b}}[{\bf q}_{s_1}(\tau), 
{\bf q}_{s_2}(\tau)]$ accumulated along $s_1$ and $s_2$ and the 
 determinant ${\cal C}_{s_1,s_2}=C_{s_1} C_{s_2}$ of the stability 
matrix corresponding to the two-particle dynamics in the 
$(d_1+d_2)-$dimensional space \cite{chaosbook}.

Our choice of initial Gaussian wave packets allows us to 
linearize the one-particle action integrals 
in ${\bf q}_{j}-{\bf r}_{i}$. We furthermore 
set ${\cal S}^{(a)}_{s_1,s_2}({\bf q}_1,{\bf x}_1;
{\bf q}_2, {\bf x}_2;t) \simeq {\cal S}^{(a)}_{s_1,s_2 }
({\bf r}_1,{\bf x}_1;{\bf r}_2, 
{\bf x}_2;t)$, keeping in mind that ${\bf r}_{1}$ and
${\bf r}_{2}$, taken as arguments of the two-particle action integrals, 
have an uncertainty ${\cal O}(\sigma)$. We then perform 
six Gaussian integrations to get
\begin{eqnarray}\label{eq:irrevtesbeforeSPA}
{\cal M}_{\rm B}(t)  &=&
   (4\pi \sigma^2)^{\frac{2 d_1+d_2}{2} }\int
 \prod_{i=1}^2 {\rm d}{\bf x}_i {\rm d}{\bf y}_i {\rm d}{\bf z}_2  \sum_{
\rm paths}
  {\cal A}_{s_1} {\cal A}_{s_2} {\cal A}_{s_3}^{\dagger} 
{\cal A}_{s_4}^{\dagger} {\cal A}_{l_1}^{\dagger} {\cal A}_{l_3}
 C_{l_2}^{{1\over 2}} C_{l_4}^{ {1\over 2} \dagger} 
\exp\left[ {\it i } \left\{ \Phi_{1}+   \Phi_{2}  +     
\Phi_{12}  \right\}\right],
\end{eqnarray}
where we wrote
${\cal A}_{s_i} = C_{s_i}^{\frac{1}{2}} \exp[- {\sigma^2 \over 2 }
({\bf p}_{s_{i}} - {\bf p}_{i})^2]$. 
Paths with odd (even) indices correspond to system 1 (2).
The semiclassical expression for ${\cal M}_{\rm B}(t)$ is obtained by enforcing
a stationary phase condition on Eq.~(\ref{eq:irrevtesbeforeSPA}), i.e. keeping
only terms which minimize the variation of the three action phases \\
\end{widetext}
 \begin{subequations}
\begin{eqnarray}
 \Phi_{1}&=& S^{(\rm f)}_{s_1}({\bf r}_1,{\bf x}_1;t) -  
S^{(\rm b)}_{l_1} ({\bf r}_1, {\bf x}_1; t) 
 \nonumber \\ &&                              
- S^{(\rm f)}_{s_3}({\bf r}_1,{\bf y}_1,t) +  S^{(\rm b)}_{l_3} ({\bf r}_1,{\bf y}_1;t),  
\label{phi1}\\
  \Phi_{2} &=& S^{(\rm f)}_{s_2}({\bf r}_2,{\bf x}_2;t)  + 
S^{(\rm b)}_{l_2}({\bf x}_2,{\bf z}_2;t) 
 \nonumber \\ &&       
                                - S^{(\rm f)}_{s_4}({\bf r}_2,{\bf y}_2;t)   - 
S^{(\rm b)}_{l_4}({\bf y}_2,{\bf z}_2;t),  \label{phi2} \\
   \Phi_{12}&=& {\cal S}^{(\rm f)}_{s_1,s_{2} } + {\cal S}^{(\rm b)}_{l_1,l_{2} } 
                                    -  {\cal S}^{(\rm f)}_{s_3,s_{4} } - 
{\cal S}^{(\rm b)}_{l_3,l_4 }.  \label{phi12}
\end{eqnarray}
\end{subequations}
The semiclassically relevant terms are identified by
path contractions. The first stationary phase approximation 
over $\Phi_{1}$ corresponds to contracting unperturbed paths with perturbed 
ones, $s_1 \simeq l_1$ and  $s_3 \simeq l_3$. This pairing 
is allowed by our assumption of a classically
weak $\Sigma_1$ \cite{caveat}.
The phase $\Phi_{1}$ is then given by the difference of
action integrals of the perturbation $\Sigma_1$ on paths $s_1$ and $s_3$,
$\Phi_{1} = \delta S_{s_1}({\bf r}_1,{\bf x}_1;t)  - 
\delta S_{s_3}({\bf r}_1,{\bf y}_1,t) $, with
  $\delta S_{s_i} =\int_{0}^{t} {\rm d}\tau \Sigma_1[{\bf q}_{s_i}(\tau)]$.
Here, ${\bf q}_{s_i}(\tau)$ lies on $s_i$ with 
${\bf q}_{s_i}(0) ={\bf r}_1$ 
and ${\bf q}_{s_1}(t)= {\bf x}_1$, ${\bf q}_{s_3}(t)= {\bf y}_1$.
A similar procedure for  
$\Phi_{2}$ requires $s_2 \simeq s_4$ and $l_2 \simeq l_4$, 
and thus ${\bf x}_2 \simeq {\bf y}_2$. These contractions lead to an exact
cancellation $\Phi_{2} = 0$, and one gets
\begin{eqnarray}\label{eq:irrevtesafterSPA}
 {\cal M}_{\rm B}(t)  &=&
   (4\pi \sigma^2)^{\frac{2 d_1+d_2}{2} }\int
 \prod_{i=1}^2 {\rm d}{\bf x}_i {\rm d}{\bf y}_j {\rm d}{\bf z}_2 \; 
\delta_\sigma({\bf x}_2-{\bf y}_2) 
 \nonumber \\ 
 \times&\!\!\! \sum &
 \vert {\cal A}_{s_1} \vert^2  \vert  {\cal A}_{s_2} \vert^2   
\vert  {\cal A}_{s_3}\vert ^2
  \vert C_{l_2}\vert
  e^{ {\it i} \left[ \delta S_{s_1}   -   \delta S_{s_3} +\Phi_{12}.
\right]}.\quad
\end{eqnarray}
Here, $\delta_\sigma({\bf x}_2-{\bf y}_2) $ restricts
the spatial integrations to
$|{\bf x}_2-{\bf y}_2|\leq \sigma$ because of the finite resolution with 
which two paths can be equated. 

The semiclassical Boltzmann Echo
(\ref{eq:irrevtesafterSPA}) is dominated by two contributions.
The first contribution is non diagonal in that all paths are 
uncorrelated. Applying the central limit theorem one has
$\left\langle \exp\left[ {\it i} \left\{ \det S_{s_1} - \det S_{s_3} + 
\Phi_{12}  \right\} \right] \right\rangle = 
\exp\left[-\left\langle  \det S^2_{s_1}\right\rangle
- \big\langle ({\cal S}^{(\rm f)}_{s_1,s_{2} })^2  \big\rangle
- \big\langle ({\cal S}^{(\rm b)}_{s_1,s_{2} })^2   \big\rangle
 \right]$, where $\langle \delta S_{s_1}^2 \rangle  =
\int_0^{t} {\rm d}\tau {\rm d}\tau' 
\langle \Sigma_1 [{\bf q}_{s_1}(\tau)] \;
\Sigma_1[{\bf q}_{s_1}(\tau')] \rangle$ and $
\big\langle ({\cal S}^{(\rm f,b)}_{s_1,s_{2} })^2  \big\rangle$ $
= \int_0^{t } {\rm d}\tau \;  {\rm d}\tau' 
\langle {\cal U}_{\rm f,b}[{\bf q}_{s_1}(\tau), {\bf q}_{s_2}(\tau)] \;
{\cal U}_{\rm f,b}[{\bf q}_{s_1}(\tau'), {\bf q}_{s_2}(\tau')]  \rangle$.
In chaotic systems, correlators typically decay exponentially fast, thus
 $\left\langle  \det S^2_{s_1}  \right\rangle \simeq  \Gamma_{\Sigma_1} \; t $  
and 
$\big\langle ({\cal S}^{(\rm f,b)}_{s_1,s_{2} })^2  \big\rangle
 \simeq \Gamma_{\rm f,b} \; t $. 
Finally using the two sum rules 
\begin{subequations}\label{sumrule}
\begin{eqnarray}
(4\pi \sigma^2)^{{{\rm d}_i \over 2}} \int {\rm d}{\bf x}_i  \sum_{s_i} \vert {\cal A}_{s_i} \vert^2 =1, \\
  \int {\rm d}{\bf x}_i  \int {\rm d}{\bf y}_i \; \delta_\sigma ( {\bf y}_i 
-{ \bf x}_i) \sum_{l_i} \vert C_{l_i} \vert =1,
\end{eqnarray}
\end{subequations}
one obtains the nondiagonal contribution 
\begin{eqnarray}
\label{eq:irrevtes1}
 {\cal M}^{(\rm nd)}_{\rm B}(t)  & \simeq & \exp\left[-\left( \Gamma_{\Sigma_1} +
\Gamma_{\rm f}+\Gamma_{\rm b} \ \right) t  \right].
\end{eqnarray}

The second contribution is diagonal, with $s_1 \simeq  s_3$ and
${\bf x}_1 \simeq {\bf y}_1$. From
Eq.~ (\ref{eq:irrevtesafterSPA}) it reads
\begin{eqnarray}\label{eq:irrevtesdiag}
 {\cal M}^{(\rm d)}_{\rm B}(t)  =&&\!\!\! \!\! 
   (4\pi \sigma^2)^{\frac{2 d_1+d_2}{2} }\int
 \prod_{i=1}^2 {\rm d}{\bf x}_i {\rm d}{\bf y}_i d{\bf z}_2  
\; \delta_\sigma({\bf x}_i-{\bf y}_i) 
 \nonumber \\ \!\!\times&&\!\!\! \!\! \!\!\! \!  \sum\!
 \vert {\cal A}_{s_1} \vert^4  
\vert  {\cal A}_{s_2}\vert ^2
 \vert C_{l_2}\vert
  e^{ {\it i} \left[ \Delta S_{s_1}+ \Delta{\cal S}_{s_1,s_2}^{({\rm f}) } + \Delta{\cal S}_{s_1,l_2} ^{({\rm b}) }
\right]},\qquad
\end{eqnarray}
where 
$\Delta S_{s_1} =  \int_{0}^{t} {\rm d}\tau \nabla_1 
\Sigma_1[{\bf q}_{s_1}(\tau)] 
\cdot [{\bf q}_{s_3}(\tau)-{\bf q}_{s_1}(\tau)]$ and 
$\Delta  {\cal S}^{({\rm f,b})}_{s_1,s_2}=  \int_{0}^{t} {\rm d}\tau
\nabla_1 {\cal U}_{{\rm f,b}}[{\bf q}_{s_1}(\tau),{\bf q}_{s_2}(\tau)] \cdot [
{\bf q}_{s_3}(\tau)-{\bf q}_{s_1}(\tau)]
$. 
We perform a change of coordinates $\int d{\bf x}_1 \sum  |{C}_{s_1}| 
 = \int d{\bf p_1}$, and use both the asymptotics
$|C_{s_1}| \propto \exp\left[-\lambda_1 t \right]$ valid for chaotic 
systems \cite{chaosbook} 
and the sum rules of Eqs.~(\ref{sumrule}) to get
 \begin{eqnarray}\label{eq:rrevtes2}
{\cal M}^{(\rm d)}_{\rm B}(t) & \simeq & 
 \alpha_1  \exp\left[-\lambda_1 t \right].
  \end{eqnarray}
Here, $\alpha_1$ is only algebraically time-dependent with
$\alpha_1(t=0) = {\cal O} (1)$. Together, diagonal (\ref{eq:rrevtes2})
and nondiagonal (\ref{eq:irrevtes1})
contributions sum up to our main result, Eq.~(\ref{eq:LEmod}).
We finally note that the long-time saturation at the inverse
Hilbert space size of system 1,
${\cal M}_{\rm B}(\infty) = N_1^{-1}$,
is obtained from Eq.~(\ref{eq:irrevtesbeforeSPA}) with the contractions
$s_1\simeq s_3$, $s_2\simeq s_4$, $l_1\simeq l_3$ and $l_2\simeq l_4$.
 
Analyzing Eq.~(\ref{eq:LEmod}), 
we first note that ${\cal M}_{\rm B}(t)$ 
depends neither on $H_2$ nor on $\Sigma_2$. 
This is so because one traces over the 
uncontrolled degrees of freedom. We stress that 
this holds even for classically strong $\Sigma_2$. Most importantly,
besides strong similarities
with the Loschmidt Echo, such as competing golden rule and Lyapunov decays
\cite{Jal01,Jac01}, the Boltzmann Echo can exhibit a $\Sigma_1$-independent 
decay given by the decoherence rates $\Gamma_{\rm f,b}$ in the limit
$\Gamma_{\Sigma_1} \ll \Gamma_{\rm f,b}$. Extending our analysis to the regime 
$\Gamma_{\Sigma_1} \ll \delta_1$, $\Gamma_{\rm f,b} \ll \delta_2$ by means of
quantum perturbation theory, we find a gaussian decay of
${\cal M}_{\rm B}(t)$, Eq.~(\ref{gaussiand}).
It is thus possible to reach either a Gaussian or an exponential, 
$\Sigma_1$-independent decay, depending on the balance between the accuracy $\Sigma_1$
with which the time-reversal operation is performed and the coupling between controlled
and uncontrolled degrees of freedom. This might explain the experimentally
observed saturation of the polarization echo as $\Sigma_1$ is reduced
\cite{levstein}, though a more precise analysis of these experiments
in the light of the results presented here is necessary.

%%%%%%%%%%%%%%%%%%%%%%%%%%%%%%%%%%%%%%%%%%%%%%%%%%%%%%%%%%%%%%%%
\begin{figure}
\rotatebox{0}{\resizebox{7.8cm}{!}{\includegraphics{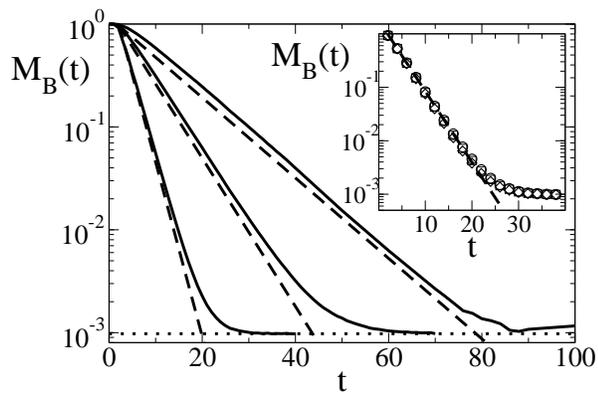}}}
\caption{\label{fig1}
Main plot: Boltzmann echo for $N=1024$, $K_1=K_2 = 10.09$, and
$\sigma_1 = 0.0018$ ($ \Gamma_{\Sigma_1} \simeq 0.09$). 
Data have been calculated from 50 different initial states.
The full lines correspond to $ \epsilon= 0 $, 0.0018 and 0.0037 (from right to left)
and the dashed lines give the predicted exponential decay of
Eq.~(\ref{eq:LEmod}), with $\Gamma_{\cal U} = 1.2\, 10^{4} \epsilon^2,
 \Gamma_{\Sigma_1}= 2.6 \, 10^{4} \sigma_1^2$, $\lambda = 1.6 \gg  \Gamma_{\cal U}, \Gamma_{\Sigma_1}$ (dashed lines have been
slightly shifted for clarity).
The dotted line gives the saturation $N^{-1}.\,$
Inset :  ${\cal M}_{\rm B}$ for $\epsilon =0.0037$,  and 
$ \sigma_1= 0.0003$ (circles; $ \Gamma_{\Sigma_1} \simeq 2.\, 10^{-3} $), $ \sigma_1= 0.0006$ (squares; $ \Gamma_{\Sigma_1} \simeq 9 .\,10^{3}$), and $0.0009$ (diamonds; $ \Gamma_{\Sigma_1} \simeq 0.02 $). 
The dashed line indicates the theoretical prediction 
${\cal M}_{\rm B}(t) = \exp[-0.3 t]$.}
\end{figure}
%%%%%%%%%%%%%%%%%%%%%%%%%%%%%%%%%%%%%%%%%%%%%%%%%%%%%%%%%%%%%%

We numerically illustrate our findings. 
We consider two coupled kicked rotators with Hamiltonian
\begin{subequations}
\label{krot}
\begin{eqnarray}
H_i & = & p_i^2 / 2 + K_i \cos(x_i) \; \sum_n \delta(t-nT),\\
{\cal U} & = & \epsilon \; \sin(x_1-x_2-0.33) \; \sum_n \delta(t-nT).
\end{eqnarray}
\end{subequations}
 We concentrate on the regime $K_i > 7$, for which the dynamics is fully chaotic with Lyapunov exponent $\lambda_i\approx\ln [K_i/2]$. The time-reversed
one-particle Hamiltonians are obtained through $K_i \rightarrow 
K_i +\sigma_i$. We here restrict ourselves to the case ${\cal U}={\cal U}_{\rm f}={\cal U}_{\rm b}$.
Both rotators are quantized on the torus with discrete 
momenta $p_n = 2 \pi n/N$, $n=1,2,...N$.
The one- and two-particle bandwidths and level spacings are given by
$B_1=2 \pi$, $\delta_1=2 \pi/N$ and $B_2 = 4 \pi$, $\delta_2 = 4 \pi/N^2$. 
For more details on the numerical
procedure, we refer the reader to Ref.~\cite{Izr}.

We first checked that ${\cal M}_{\rm B}(t)$ is independent of $K_2$ (as long as
system 2 remains chaotic) and $\sigma_2$, and therefore set 
$K_2=K_1$, $\sigma_2=0$. The main panel in Fig.~\ref{fig1} shows that for 
$B_1 > \Gamma_{\Sigma_1}  > \delta_1$,  $B_2 > \Gamma_{\cal U}  > \delta_2$,
Eq.~(\ref{eq:LEmod}) is satisfied. Additionally, the inset of
Fig.~\ref{fig1} illustrates that when 
$\Gamma_{\Sigma_1} \ll 2 \Gamma_{\cal U}$,
the observed decay is only sensitive to ${\cal U}$ , 
and one effectively obtains a $\Sigma_1$-independent decay. Further unshown 
data confirm the existence of the Lyapunov decay 
[second term in Eq.~(\ref{eq:LEmod})]. All our 
numerical results thus confirm the validity of Eq.~(\ref{eq:LEmod}).
   
In conclusion we propose to analyze echo experiments in the light of
the {\it Boltzmann echo} of Eq.~(\ref{irrevtest}) and (\ref{hamiltonians}). 
Our semiclassical and RMT analysis 
showed that the decay of ${\cal M}_{\rm B}(t)$ saturates at a finite value even
when the time-reversal operation is performed with infinite accuracy. 
Further work should attempt to connect these results with
echo experiments \cite{nmr,kurnit,davidson,nakamura,levstein}.

One of us (CP) acknowledges the support of 
the Swiss National Science Foundation.

\end{document}